\documentclass[prmaterials,reprint,superscriptaddress]{revtex4-1}

\usepackage[T1]{fontenc}
\usepackage[utf8]{inputenc}
\usepackage{lmodern,times,amsmath,amssymb,amsfonts,textcomp}
\usepackage[colorlinks=true,allcolors=blue]{hyperref}
\usepackage{graphicx}

\newcommand{\matfam}{YCu$_{3}$(OH)$_{6}$O$_{x}$Cl$_{3-x}$ ($x=0,1/3$)}
\newcommand{\mata}{YCu$_{3}$(OH)$_{6}$Cl$_{3}$}
\newcommand{\matb}{Y$_{3}$Cu$_{9}$(OH)$_{18}$OCl$_{8}$}
\newcommand{\YYY}{Y$_{3}$Cu$_{9}$(OH)$_{19}$Cl$_{8}$}
\newcommand{\HS}{ZnCu$_{3}$(OH)$_{6}$Cl$_{2}$}
\newcommand{\LiHS}{ZnLi$_{x}$Cu$_{3}$(OH)$_{6}$Cl$_{2}$}
\newcommand{\GaHS}{Ga$_{x}$Cu$_{4-x}$(OD)$_{6}$Cl$_{2}$}
\newcommand{\CuBaVOH}{Cu$_{3}$Ba(VO$_{5}$H)$_{2}$}

\begin{document}

\title{Local study of the insulating quantum kagome antiferromagnets \matfam}

\author{Quentin Barth\'{e}lemy}
\affiliation{Laboratoire de Physique des Solides, UMR 8502, Univ. Paris-Sud, Universit\'{e} Paris-Saclay, 91405 Orsay, France}
\author{Pascal Puphal}
\affiliation{Physikalisches Institut, Goethe-Universit\"{a}t Frankfurt, 60438 Frankfurt am Main, Germany}
\affiliation{Laboratory for Multiscale Materials Experiments, Paul Scherrer Institute, 5232 Villigen, Switzerland}
\author{Katharina M. Zoch}
\affiliation{Physikalisches Institut, Goethe-Universit\"{a}t Frankfurt, 60438 Frankfurt am Main, Germany}
\author{Cornelius Krellner}
\affiliation{Physikalisches Institut, Goethe-Universit\"{a}t Frankfurt, 60438 Frankfurt am Main, Germany}
\author{Hubertus Luetkens}
\affiliation{Laboratory for Muon-Spin Spectroscopy, Paul Scherrer Institut, 5232 Villigen, Switzerland}
\author{Christopher Baines}
\affiliation{Laboratory for Muon-Spin Spectroscopy, Paul Scherrer Institut, 5232 Villigen, Switzerland}
\author{Denis Sheptyakov}
\affiliation{Laboratory for Neutron Scattering and Imaging, Paul Scherrer Institute, 5232 Villigen, Switzerland}
\author{Edwin Kermarrec}
\affiliation{Laboratoire de Physique des Solides, UMR 8502, Univ. Paris-Sud, Universit\'{e} Paris-Saclay, 91405 Orsay, France}
\author{Philippe Mendels}
\affiliation{Laboratoire de Physique des Solides, UMR 8502, Univ. Paris-Sud, Universit\'{e} Paris-Saclay, 91405 Orsay, France}
\author{Fabrice Bert}
\affiliation{Laboratoire de Physique des Solides, UMR 8502, Univ. Paris-Sud, Universit\'{e} Paris-Saclay, 91405 Orsay, France}

\date{\today}

\begin{abstract}
The quantum kagome antiferromagnets \matfam\  are produced using a unified solid state synthesis route for polycrystalline samples. From structural refinements based on neutron diffraction data, we clarify the structure of the \matb\ ($x=1/3$) compound and provide a revised chemical formula. We use muon spin relaxation, as a local probe of magnetism, to investigate the exotic low temperature properties in the two compounds. In agreement with the low temperature neutron diffraction data, we find no evidence for long range ordering in both materials but they exhibit distinct ground states: while disordered static magnetism develops in the $x=0$ compound, we conclude on the stabilization of a quantum spin liquid in the $x=1/3$ one, since the local fields remain fully dynamical. Our findings are in contrast to previous reports based on thermodynamical measurements only. We then discuss their origin on the basis of structural details and specific heat measurements. In particular, the $x=1/3$ compound appears to realize an original spatially anisotropic kagome model.
\end{abstract}

\maketitle

\section{Introduction}

Since Anderson's founding ideas~\cite{Anderson1973,Anderson1987}, the prospect of generating high-$T_{\mathrm{c}}$ superconductivity by doping a Mott insulator in the Quantum Spin Liquid (QSL) state remains a promising route that many are following theoretically~\cite{Senthil2005,Lee2006,Lee2008}. Similar effort can now be initiated on the experimental side, enabled by the syntheses of materials presenting essential attributes of a QSL: no spontaneous symmetry breaking down to the lowest temperatures and the presence of a continuum of fractional excitations.\\
\indent Over the last ten years, \HS\ polymorphs herbertsmithite and kapellasite have been recognized as model realizations of QSL among two-dimensional non-organic copper-oxide minerals~\cite{Mendels2007,Olariu2008,Han2012,Fak2012,Jeschke2013,Kermarrec2014}. In these two atacamites, Cu$^{2+}$ spins $S=1/2$ decorate a geometrically perfect kagome lattice. The main structural difference arises from the position of zinc cations: in between the kagome planes for herbertsmithite and within the planes for kapellasite. The superexchange paths are substantially different in both compounds providing two different archetypal frustated models. In herbertsmithite, the dominant term is the antiferromagnetic exchange between nearest neighbours. In kapellasite, frustration occurs from the competition of ferromagnetic $J_{1}$ and antiferromagnetic $J_{\mathrm{d}}$ exchanges between nearest neighbors and along the diagonals of the hexagons respectively. The QSL ground states that originate in both cases from the low dimensionality and the set of frustrated interactions appears as a good playground for doping in order to reach superconductivity. Besides, adding carriers in a kagome Mott insulator is interesting in itself, considering the inherent hexagonal symmetry: for a $n=4/3$ filling, the band structure is predicted to feature Dirac cones at the Fermi level~\cite{Mazin2014,Guterding2016}.\\
\indent On the chemistry side, it has proven to be challenging to directly synthesize an electron-doped compound from the formula \HS. A first attempt was the successful topochemical intercalation of a nominal $n$-type dopant: \LiHS\ has been synthesized for $x$ up to $1.8$ ($n=8/5$)~\cite{Kelly2016}. More recently, a successful substitution of Zn$^{2+}$ with a trivalent cation has been reported: \GaHS\ has been synthesized for $x$ up to $0.8$ ($n=4/3.2$)~\cite{Puphal2019}. Unfortunately, doping is not effective in these systems in the sense that no metallic conductivity has been observed. In Zn-Cu hydroxyl halides kagome materials it has been proposed that additional charges do not delocalize but form self-trapped Cu$^{+}$ polarons~\cite{Liu2018}. The bright side of these attempts is that a wealth of new materials that are very close to their parents herbertsmithite and kapellasite is being generated, which may show less magnetic dilution on the lattice and provide also new models.\\
\indent This article focuses on two fortuitous outcomes in trying to replace Zn$^{2+}$ with trivalent Y$^{3+}$. This led to the insulators \matfam\ which display a kapellasite-like structure and no sign of Cu/Y mixing from single crystal x-ray refinements, in contrast to the previously reported Zn-Cu systems~\cite{Sun2016,Puphal2017,Pustogow2017}. In the $x=0$ compound, the kagome lattice is perfect and the charge is balanced by additional anions. In the $x=1/3$ compound, initially identified as the stoichiometric \YYY, the kagome lattice is slightly distorted and the excess charge ($n=8/9$) remains localized. Susceptibility measurements have displayed a rather large antiferromagnetic Weiss temperature close to $\theta\sim-100$~K for both compounds and no sign of magnetic transition at $|\theta|$, pointing to magnetic frustration and/or weak interlayer coupling. At lower temperature, a peak in the specific heat occurs around $2$~K for $x=1/3$ which represents a small entropy gain of about $0.1R\ln{(2)}$, while only a broad maximum is detected in the $5$~T magnetization in this range~\cite{Puphal2017}. For $x=0$, the magnetic dc susceptibility steeply increases below $\sim15$~K and becomes hysteretic below $\sim6.5$~K although ac susceptibility hardly depends on temperature nor on frequency below $\sim15$~K~\cite{Sun2016}. From these thermodynamic data, it was suggested that a magnetic transition occurs in the $x=1/3$ compound at around $2$~K while the $x=0$ compound may realize a spin liquid ground state. Motivated by these former intriguing results, we investigated further the two closely related materials.\\
\indent A new unified synthesis route is described to produce polycrystalline samples with composition $x=0$ or $x=1/3$ by varying the reaction temperature. From neutron diffraction data at $1.6$~K, we refine the crystal structure parameters. We further find no evidence for long-range magnetic order, neither for $x=0$ nor for $x=1/3$. Finally, to reveal unambiguously the nature of the ground states, we have performed a Muon Spin Relaxation ($\mu$SR) study  which leads to the identification of two distinct magnetic states, respectively static and dynamical for $x=0$ and $x=1/3$. The occurrence of these two ground states is discussed in relation to the crystal structures of both compounds and a comparison of their specific heat.

\section{Powder synthesis}

The syntheses of the \matfam\ compounds were previously described in References~\cite{Sun2016,Puphal2017}. Despite their chemical similarity, no transition in the synthesis from one material to the other was found so far, as the $x=0$ compound could only be produced from a non-hydrothermal flux method while crystals of composition $x=1/3$ were obtained with a hydrothermal route. Here, we are able to synthesize both compounds from a solid state reaction in a teflon lined vessel by sintering pellets of the stoichiometric mixture of YCl$_{3}$\textbullet$6$H$_{2}$O with malachite Cu$_{2}$(OH)$_{2}$CO$_{3}$. Applying this method, the $x=0$ compound (P$\bar{3}$m$1$) forms at temperatures below $180$~\textdegree~C while the $x=1/3$ compound (R$\bar{3}$) forms above $190$~\textdegree~C. This transition occurs due to a strong release of water from YCl$_{3}$\textbullet$6$H$_{2}$O leading to the decay of the $x=0$ composition to the $x=1/3$ one, below the water-free atmosphere decaying temperature of $350$~\textdegree~C~\cite{Sun2016}. For temperatures above $220$~\textdegree~C, the clinoatacamite impurity amount increases. In both cases, pellets of bright blue powder can be collected without the neccessity of filtration.\\
\indent We note, as in Reference \cite{Sun2016}, that the $x=0$ compound decays once it comes into contact with water. Repeated measurements expose it to moisture and a gradual increase of the amount of the $x=1/3$ compound can be observed. Because of the decay in free water, deuteration of the $x=0$ compound was not successful. A deuteration of the pre-reactants as well as of the reacted phase always leads to a phase mixture containing clinoatacamite and the $x=1/3$ compound. A deuterated powdered sample of the $x=1/3$ compound, with a composition Y$_{3}$Cu$_{9}$(OD)$_{18}$OCl$_{8}$, was nonetheless prepared for a neutron powder diffraction study using $1.1$~g of dried YCl$_{3}$ salt, $0.787$~g of CuO and $4$~mL of D$_{2}$O. This precursor mixture was heated to $220$~\textdegree~C for seven days in a Parr reaction vessel.

\section{Low-temperature neutron diffraction}

\begin{figure}\centering
\includegraphics[width=\linewidth]{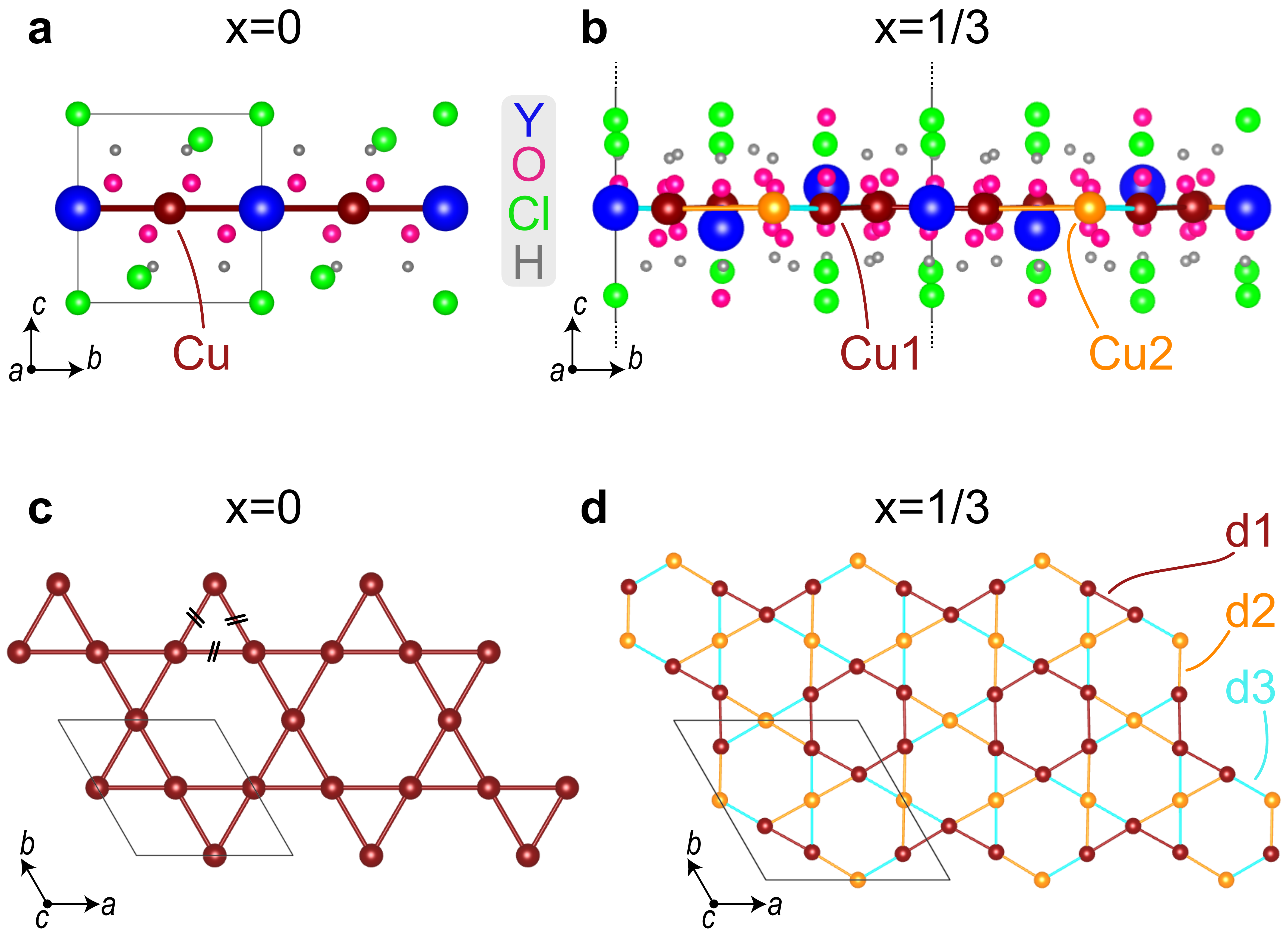}
\caption{\label{fig1}(a) Crystal structure of YCu$_{3}$(OH)$_{6}$Cl$_{3}$ ($x=0$ compound) projected along $\mathbf{a}$. (b) View of one isolated kagome plane in Y$_{3}$Cu$_{9}$(OH)$_{18}$OCl$_{8}$ ($x=1/3$ compound) projected along $\mathbf{a}$, showing the buckling of the Cu-Y layer. We assume that the H atoms are located at the same positions as the D atom ones, obtained from neutron diffraction refinements and shown in Table~\ref{tab1}. Note that the orientation the $\mathbf{a}$ direction with respect to the kagome plane is different for both compounds. (c) and (d) show only the magnetic ions Cu$^{2+}$ in the $\mathbf{a}-\mathbf{b}$ plane for the $x=0$ and $x=1/3$ compounds respectively. The $x=0$ compound exhibits a geometrically perfect kagome lattice whereas the $x=1/3$ compound features an anisotropic kagome lattice with two copper sites Cu$1$, Cu$2$, and three distinct lengths listed as $d1$, $d2$ and $d3$.}
\end{figure}

\begin{figure}\centering
\includegraphics[width=\linewidth]{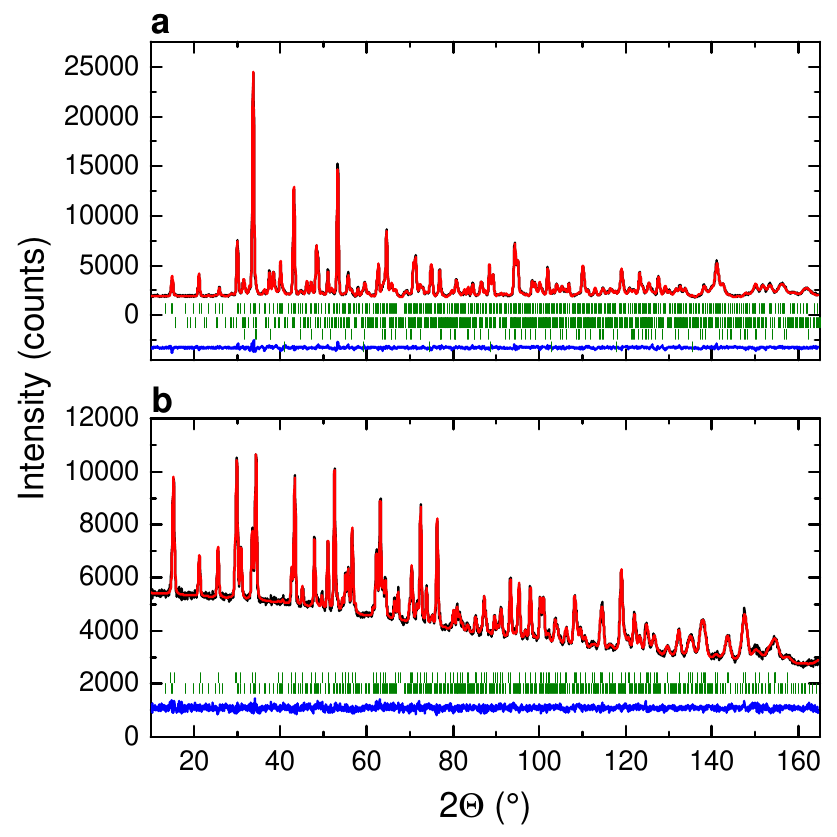}
\caption{\label{fig2}Rietveld refinement plots for (a) Y$_{3}$Cu$_{9}$(OD)$_{18}$OCl$_{8}$ (deuterated $x=1/3$ compound) and (b) YCu$_{3}$(OH)$_{6}$Cl$_{3}$ ($x=0$ compound), based on neutron powder diffraction data measured at $T=1.6$~K with a wavelength $\lambda=1.494$~\AA. The observed intensity (black), calculated profile (red), and difference curve (blue) are shown. The rows of green ticks correspond to the calculated diffraction peak positions of the involved phases. In (a), from top to bottom, they are: Y$_{3}$Cu$_{9}$(OD)$_{18}$OCl$_{8}$, Cu$_{2}$(OD)$_{3}$Cl and CuO. In (b), from top to bottom, they are: YCu$_{3}$(OH)$_{6}$Cl$_{3}$ and Y$_{3}$Cu$_{9}$(OH)$_{18}$OCl$_{8}$.}
\end{figure}

\begin{table*}\centering
\begin{tabular}{|l l cc||l l c c|}
\hline
\multicolumn{4}{|c||}{YCu$_{3}$(OH)$_{6}$Cl$_{3}$ ($x=0$)} & \multicolumn{4}{|c|}{Y$_{3}$Cu$_{9}$(OD)$_{18}$OCl$_{8}$ ($x=1/3$)}\\
\multicolumn{4}{|c||}{Space group: P$\overline{3}$m$1$ (No. 164), $Z=1$} & \multicolumn{4}{|c|}{Space group: R$\overline{3}$ (No. 148), $Z=3$}\\
\hline
$a,\AA$&&&6.74022 (15)&\hspace{3mm}$a,\AA$&&&11.5326 (3)\\
$c,\AA$&&&5.59929 (13)&\hspace{3mm}$c,\AA$&&&17.1437 (4)\\
$\rm V,\AA^3$&&&220.299 (9)\hspace{3mm}&\hspace{3mm}$\rm V,\AA^3$&&&1974.64 (8)\\
\hline
Y&$1b\,(0,0,1/2)$&$\rm B,\,\AA^2$&0.47 (5)&\hspace{3mm}Y1&$6c\,(0,0,z)$&$z$&0.1290 (4)\\
&&&&\hspace{3mm}Y2& $3b\,(0,0,1/2)$ &$\rm B(Y1,Y2),\,\AA^{2}$&0.52 (8)\\
\hline
Cu&$3f\,(1/2,0,1/2)$&$\rm B,\,\AA^2$&0.42 (2)&\hspace{3mm}Cu1&$18f\,(x,y,z)$&$x$&0.6621 (5)\\
&&&&\hspace{3mm}&&$y$&0.8257 (6)\\
&&&&\hspace{3mm}&&$z$&0.5018 (3)\\
&&&&\hspace{3mm}Cu2&$9d\,(1/2,0,1/2)$&$\rm B(Cu1,Cu2),\,\AA^{2}$&0.45 (3)\\
\hline
Cl1&$2d\,(1/3,2/3,z)$&$z$&0.8646 (4)&\hspace{3mm}Cl1&$18f\,(x,y,z)$&$x$&0.6707 (5)\\
&&&&\hspace{3mm}&&$y$&0.0006 (5)\\
&&&&\hspace{3mm}&&$z$&0.28389 (14)\\
Cl2&$1a\,(0,0,0)$&$\rm B(Cl1,Cl2),\AA^2$&0.45 (2)&\hspace{3mm}Cl2&$6c\,(0,0,z)$&$z$&0.3385 (3)\\
&&&&\hspace{3mm}&&$\rm B(Cl1,Cl2),\AA^2$&0.49 (3)\\
\hline
&&&&\hspace{3mm}O1&$3a\,(0,0,0)$&$\rm B(O1,O2,O3,O4),\AA^2$&0.36 (3)\\
O&$6i\,(x,-x,z)$&$x$&0.19095 (14)&\hspace{3mm}O2&$18f\,(x,y,z)$&$x$&0.8132 (6)\\
&&$z$&0.3658 (3)&\hspace{3mm}&&$y$&0.8032 (7)\\
&&$\rm B,\AA^2$&0.59 (3)&\hspace{3mm}&&$z$&0.5435 (3)\\
&&&&\hspace{3mm}O3&$18f (x,y,z)$&$x$&0.5329 (6)\\
&&&&\hspace{3mm}&&$y$&0.6671 (10)\\
&&&&\hspace{3mm}&&$z$&0.5559 (3)\\
&&&&\hspace{3mm}O4&$18f\,(x,y,z)$&$x$&0.5045 (7)\\
&&&&\hspace{3mm}&&$y$&0.8395 (7)\\
&&&&\hspace{3mm}&&$z$&0.4637 (3)\\
\hline
H&$6i\,(x,-x,z)$&$x$&0.2039 (3)&\hspace{3mm}D2&$18f\,(x,y,z)$&$x$&0.8040 (7)\\
&&$z$&0.1914 (6)&\hspace{3mm}&&$y$&0.7954 (7)\\
&&$\rm B,\AA^2$&1.58 (5)&\hspace{3mm}&&$z$&0.5983 (4)\\
&&&&\hspace{3mm}D3&$18f\,(x,y,z)$&$x$&0.5632 (6)\\
&&&&\hspace{3mm}&&$y$&0.6596 (10)\\
&&&&\hspace{3mm}&&$z$&0.6073 (4)\\
&&&&\hspace{3mm}D4&$18f\,(x,y,z)$&$x$&0.4969 (8)\\
&&&&\hspace{3mm}&&$y$&0.8293 (9)\\
&&&&\hspace{3mm}&&$z$&0.4087 (4)\\
&&&&\hspace{3mm}&&$\rm B(D2,D3,D4),\AA^2$&1.83 (6)\\
\hline
\end{tabular}
\caption{\label{tab1}Refined crystal structure parameters of YCu$_{3}$(OH)$_{6}$Cl$_{3}$ ($x=0$ compound) and Y$_{3}$Cu$_{9}$(OD)$_{18}$OCl$_{8}$ (deuterated $x=1/3$ compound) from neutron powder diffraction ($\lambda=1.494$~\AA) at $1.6$~K. Atomic displacement parameters $B$ were constrained to equality for identical atoms.}
\end{table*}

The two compounds \matfam\ have closely related, yet slightly different, crystal structures as illustrated in Figure~\ref{fig1}. As previously described in References~\cite{Sun2016,Puphal2017}, they both crystallize in the kapellasite-type structure with a AA stacking of the kagome layers and the Y$^{3+}$ located at, or close to, the centers of the stars, the magnetic planes being well separated by Cl$^{-}$. The x-ray diffraction patterns of the two materials look quite similar, implying only minor differences in the structures. At room temperature, the indexing reveals a trigonal unit cell with $a_{x=0}\sim6.74$~\AA\ and $c_{x=0}\sim5.99$~\AA\ for $x=0$ while the x-ray powder pattern of $x=1/3$ indicates a superstructure with $a_{x=1/3}\sim\sqrt{3}a_{x=0}$ and $c_{x=1/3}\sim3c_{x=0}$ and thus a nine times larger unit cell volume.\\
\indent Neutron powder diffraction experiments were carried out with the HRPT high resolution powder diffractometer at SINQ spallation source~\cite{Fischer2000}. Approximately $1$~g of each sample were enclosed into vanadium cans with inner diameters of $\sim6$~mm and the measurement was performed at $1.6$~K and at room temperature with a wavelength $\lambda=1.494$~\AA. The seeming differences in the neutron diffraction patterns of the two materials (see Figure~\ref{fig2}, where the data collected at $T=1.6$~K are shown) are mainly due to the fact that $x=0$ only contains natural hydrogen, while $x=1/3$ is fully deuterated. The strong differences in the coherent scattering lengths and incoherent scattering cross-sections for natural hydrogen and deuterium are conditioning these drastic differences in the peak intensities and in the backgrounds of the data from two compounds. As a result of deuteration, the $x=1/3$ sample contains substantial amounts of two impurity phases: clinoatacamite Cu$_{2}$(OD)$_{3}$Cl ($\sim7.3$~wt\%) and CuO ($\sim11.3$~wt\%), with the main phase amounting to $\sim81.0$~wt\%. The only impurity present for the $x=0$ sample is actually the $x=1/3$ composition ($\sim7.0$~wt\%), to which it decays with moisture.\\
\indent For the two samples, the refined parameters of the crystal structure are given in Table \ref{tab1}. In both cases, we can discard any possible structural transition from room temperature down to $1.6$~K and we did not observe additional Bragg peaks which would have been attributed to long-range magnetic order. For $x=0$, we confirm the P$\bar{3}$m$1$ structure but we reject any statistical significance of using a disordered model for the Y$^{3+}$ site as was assumed in Reference~\cite{Sun2016}. The latter result, which was obtained in a single crystal refinement, might rather be interpreted as a sign of the slow decay towards the $x=1/3$ composition. For the $x=1/3$ compound, we make a major correction to our previously reported results~\cite{Puphal2017} where an additional disordered hydrogen ion was used. Its absence in the structure is now confirmed. Thus, the only chemical difference between these two compounds is a substitution of $1/9^{\mathrm{th}}$ of all Cl$^{-}$ by O$^{2-}$, that leads to the composition Y$_{3}$Cu$_{9}$(OH)$_{18}$OCl$_{8}$. In our description of Wyckoff positions for $x=1/3$, these oxygens O$1$ are positioned in the $3$a site. In terms of crystal structure, this substitution occurs in an ordered manner and yields a superstructure unit cell with a different symmetry.\\
\indent Unlike in the $x=0$ compound, where all the kagome units are identical and ideally flat with nearest neighboring Cu$^{2+}$ separated by a unique distance of $a_{x=0}/2\sim3.3701(1)$~\AA, two slightly different types of kagome units are being observed in the $x=1/3$ compound (as shown in Figure~\ref{fig1}(c,d)). One third of all kagome units have the internal hexagon made by Cu$1$ coppers only. The Cu$1$-Cu$1$ distance is unique, it is $\sim3.376(1)$~\AA\ ($d1$ in Figure~\ref{fig1}(d)). The other two thirds have the internal hexagon formed by Cu$1$ and Cu$2$ coppers. Cu$1$-Cu$2$ bonds sequentially alternate between $\sim3.362(8)$~\AA\ and $\sim3.252(6)$~\AA\ ($d2$ and $d3$ in Figure~\ref{fig1}(d)). The magnetic layers get very slightly buckled and the maximal ``thickness'' along $\mathbf{c}$ amounts to $\sim0.06(1)$~\AA. Furthermore, two thirds of all ytrium ions (those labelled Y$1$ in Table~\ref{tab1}) do depart from their ideal position in the center of the stars and are shifted by $\sim0.646(7)$~\AA\ along $\mathbf{c}$ towards the substitutional oxygen ions.

\section{Muon spin relaxation}

\begin{figure}\centering
\includegraphics[width=\linewidth]{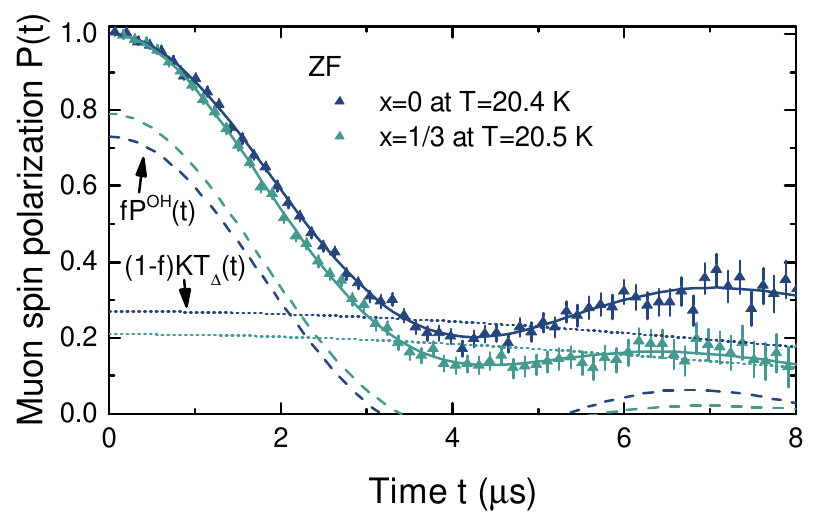}
\caption{\label{fig3}Zero field high temperature muon spin polarization in both compounds. The solid lines are fits to Equation~\ref{eq1}. The dashed and dotted lines respectively indicate the fractions of the signals associated to the two kinds of stopping sites.}
\end{figure}

To further explore the low temperature magnetic properties of these two QSL candidates, we performed $\mu$SR experiments on the GPS and LTF instruments at the PSI muon facility, in zero field (ZF) and longitudinal applied field (LF) geometries down to $20$~mK. Fully spin polarized positive muons are implanted into the sample where they stop at the most electronegative locations at time zero. Local fields then dictate the muon spin evolution during a mean lifetime of about $2.2$~$\mu$s, before the muon decays into a positron. Positrons are preferentially emitted in the final direction of the muon spins, allowing to measure the time evolution of the muon spin polarization~\cite{Bert2014,Blundell1999}. $\mu$SR is a tool of choice to investigate the elusive magnetic ground states of the $x=0$ and $x=1/3$ compounds since it directly probes the local moments and their fluctuations with a unique accuracy.\\
\indent The evolution of the ZF muon spin polarization in the temperature independent paramagnetic regime ($T>20$~K) is depicted in Figure~\ref{fig3}. For both materials, due to the motional narrowing of the fast fluctuating electronic spins, the $\mu$SR signal is dominated by the quasi-static random magnetism of the non-interacting nuclear spins. We observe a slow oscillation, a hallmark of muons forming $\mu$-OH complexes~\cite{Schenck1971,Lord2000}, with the corresponding polarization:
\begin{equation}
\begin{split}
P^{\mathrm{OH}}(t)=\exp{\left(-\frac{\Delta_{\mathrm{OH}}^{2}t^{2}}{2}\right)}\left[\frac{1}{6}+\frac{1}{3}\cos{\left(\frac{\omega_{\mathrm{OH}}t}{2}\right)}\right.\\\left.+\frac{1}{6}\cos{(\omega_{\mathrm{OH}}t)}+\frac{1}{3}\cos{\left(\frac{3\omega_{\mathrm{OH}}t}{2}\right)}\right],
\end{split}
\end{equation}
where the pulsation
\begin{equation}
\omega_{\mathrm{OH}}=\frac{\hbar\mu_{0}\gamma_{\mu}\gamma_{\mathrm{H}}}{4\pi d^{3}},
\end{equation}
is related to the muon-hydrogen dipolar interaction and therefore to the $\mu$-H distance $d$ through the muon $\gamma_{\mu}=851.616$~Mrad/s/T and proton $\gamma_{\mathrm{H}}=267.513$~Mrad/s/T gyromagnetic ratios. We added the gaussian broadening $\exp{(-\Delta_{\mathrm{OH}}^{2}t^{2}/2)}$ to account for the influence of the surrounding random nuclear fields. As in other hydroxyl halide or hydroxyl sulfate kagome materials~\cite{Mendels2007,Fak2012,Gomilsek2016} the polarization in the paramagnetic limit for both compounds can be fitted to the following expression:
\begin{equation}\label{eq1}
P_{\mathrm{ZF}}^{\mathrm{HT}}(t)=fP^{\mathrm{OH}}(t)+(1-f)\mathrm{KT}_{\Delta}(t),
\end{equation}
where a fraction $f$ of muons forms $\mu$-OH complexes while the remaining fraction - likely muons stopping close to Cl$^{-}$ and also O$^{2-}$ in the $x=1/3$ compound - experiences a nearly static gaussian nuclear field distribution of width $\Delta$ yielding the usual $\mathrm{KT}_{\Delta}(t)$ Kubo-Toyabe relaxation~\cite{Lee1999,Yaouanc2010}. As expected from their similar structure, we obtain rather similar parameters for both materials which are summarized in Table~\ref{tab2}.

\begin{table}\centering
\begin{tabular}{|c|c|c|}
\hline
 & $x=0$ & $x=1/3$\\
\hline
$f$ (\%) & $72.81\pm0.91$ & $79.03\pm2.00$\\
\hline
$\omega_{\mathrm{OH}}$ ($\mu$s$^{-1}$) & $0.59\pm0.01$ & $0.55\pm0.02$\\
\hline
$d$ (\AA) & $1.60 \pm0.01$ & $1.63 \pm 0.02$ \\
\hline
$\Delta$ ($\mu$s$^{-1}$) & $0.08\pm0.01$ & $0.09\pm0.02$\\
\hline
$\Delta_{\mathrm{OH}}$ ($\mu$s$^{-1}$) & $0.21\pm0.02$ & $0.29\pm0.03$\\
\hline
\end{tabular}
\caption{\label{tab2}Parameters obtained from the fits of the ZF high-temperature data displayed in Figure~\ref{fig3} to Equation~\ref{eq1}. As commonly observed, we find that the $\mu$-H distance $d$ is slightly elongated as compared to the H-H distance of about $1.515$~\AA\ in water molecules.}
\end{table}

\begin{figure}\centering
\includegraphics[width=\linewidth]{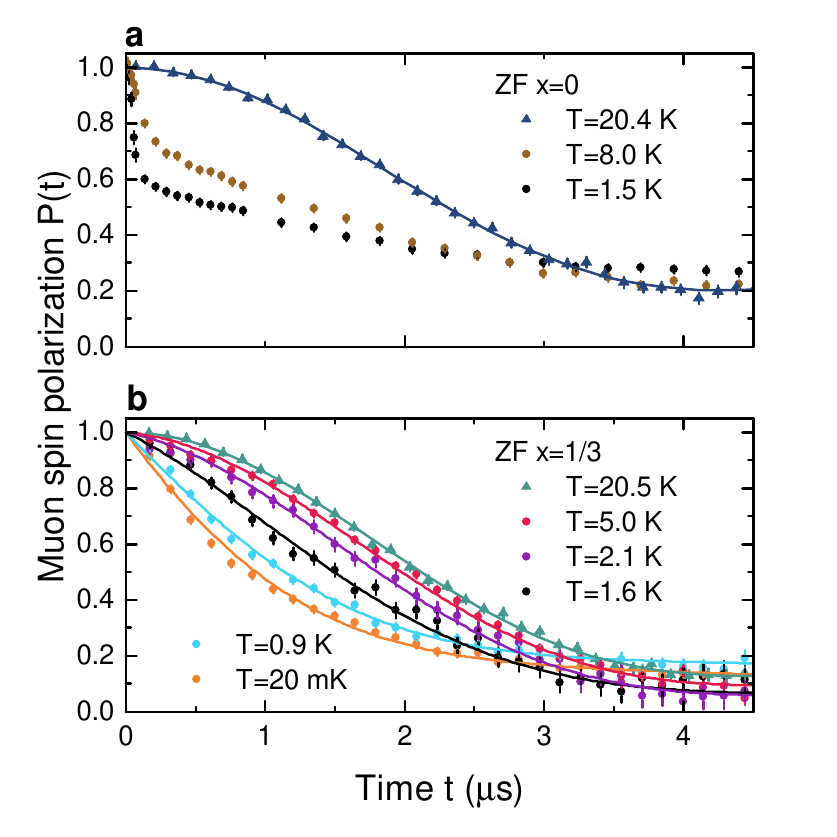}
\caption{\label{fig4}(a) Zero field muon spin polarization in the $x=0$ compound in the paramagnetic state ($20.3$~K) and at lower temperatures ($8.0$~K and $1.5$~K). The solid line is the same fit as in Figure~\ref{fig3}. (b) Zero field muon spin polarization in the $x=1/3$ compound in the paramagnetic state ($20.5$~K) and at lower temperatures (from $5.0$~K to $20$~mK). The solid line associated to $T=20.5$~K is the same fit as in Figure~\ref{fig3}. The other solid lines are fits to Equation~\ref{eq3}.}
\end{figure}

\begin{figure}\centering
\includegraphics[width=\linewidth]{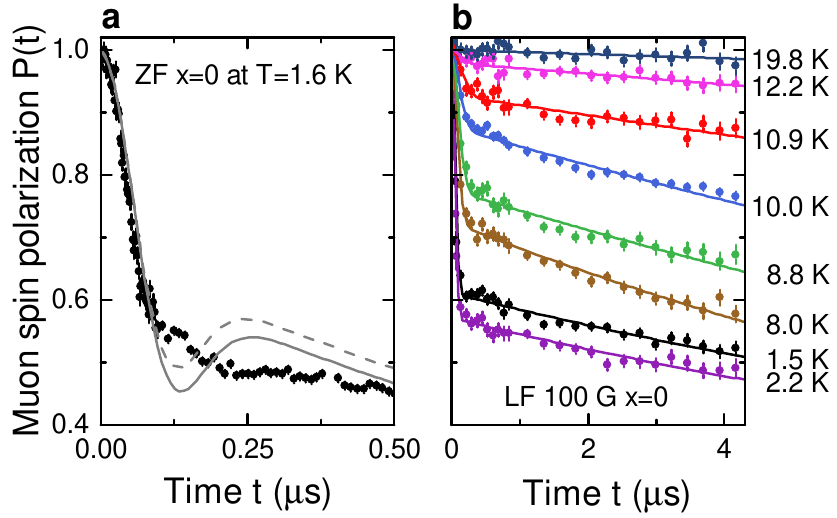}
\caption{\label{fig5}(a) A high-statistics detailed view of the early time zero field spin polarization in the $x=0$ compound at $1.6$~K demonstrating the absence of spontaneous oscillations within the accuracy of the experiment. The gray solid line is a comparison to a gaussian Kubo-Toyabe relaxation plus a dynamical component. The gray dashed line is a simulation to show how behaves the latter with a $10$~mT longitudinal applied field. (b) Temperature evolution of the muon spin polarization in the $x=0$ compound when a longitudinal field of $10$~mT is applied, from $19.8$~K to $1.5$~K. The solid lines are fits to Equation~\ref{eq2}.}
\end{figure}

\begin{figure}\centering
\includegraphics[width=\linewidth]{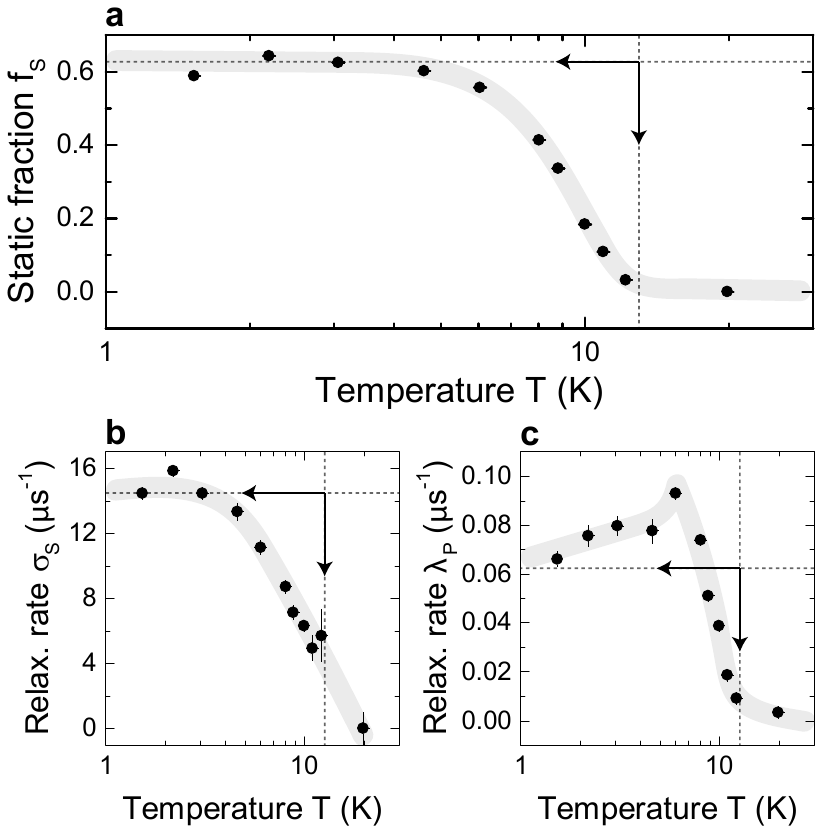}
\caption{\label{fig6}Temperature evolution of the Equation~\ref{eq2} parameters for fits on the $x=0$ data in a longitudinal applied field of $10$~mT. The light gray lines are guides to the eye, downward arrows indicate the position of the anomaly in dc susceptibility and leftward arrows indicate the estimated values of the parameters as $T\rightarrow0$. (a) The static fraction $f_{\mathrm{S}}$, that characterizes the volumic frozen fraction in the sample. (b) The fast relaxation rate $\sigma_{\mathrm{S}}$, due to frozen moments. (c) The slow relaxation rate $\lambda_{\mathrm{P}}$, due to the remaining paramagnetic moments.}
\end{figure}

\indent As depicted in Figure~\ref{fig4}, a significant difference appears between the two $\mu$SR signals when the temperature is lowered. The early time relaxation in the $x=0$ compound increases drastically, which usually indicates a freezing of the internal fields. On the opposite, the moderate changes of the relaxation in the $x=1/3$ compound all the way down to the lowest temperature $T=20$~mK rather supports the idea of a dynamical ground state with no static internal field, the attribute of a QSL.\\
\indent The fast early time relaxation in the $x=0$ compound is shown in Figure~\ref{fig5}(a). The development of static long range order would be typically characterized by the appearance of spontaneous oscillations in the first $\mu$s. The absence of such oscillations within the accuracy of our measurements points rather to a disordered magnetic state. In order to estimate the static field distribution responsible for this fast relaxation, we compare the experimental data to a standard gaussian Kubo-Toyabe relaxation. The initial gaussian shape is well reproduced for a distribution width of $\sim 15$~mT. For comparison, the dipolar field at the oxygen site generated by a $1\mu_{\mathrm{B}}$ magnetic moment on the closest copper ion is $\sim0.17(6)$~T. Despite the heavy dependence on the unknown muon stopping site, the former estimation suggests a strongly reduced frozen moment
in the low temperature phase, of the order of $\sim0.1\mu_{\mathrm{B}}$.\\
\indent To further track the development of the static magnetic volume fraction $f_{\mathrm{S}}$ in the $x=0$ compound as a function of temperature, we applied a longitudinal field $B_{\mathrm{LF}}=10$~mT to probe the electronic spin contribution in isolation. Some of these measurements are depicted in Figure~\ref{fig5}(b) at various temperatures. The value of $B_{\mathrm{LF}}$ is chosen to be large enough compared to $\omega_{\mathrm{OH}}/\gamma_{\mu}\sim0.7$~mT, to decouple the quasi-static nuclear relaxation (see the negligible relaxation at $T=19.8$~K in Figure~\ref{fig5}(b)), but smaller than the static internal fields $\sim15$~mT so that the corresponding relaxation is hardly affected (see the simulation in Figure~\ref{fig5}(a)). We can note that the relaxation curve at $1.5$~K is above the one at $2.2$~K, showing the development of a distinctive one-third tail, expected for static magnetism in a non-oriented polycrystalline sample. Additionally, the slow relaxation at long time in these longitudinal field data demonstrates the persistence of a dynamical volume fraction. Thus, the muon spin polarization in Figure~\ref{fig5}(b) was fitted to:
\begin{equation}\label{eq2}
P_{\mathrm{LF}}(t)=f_{\mathrm{S}}\left[\frac{2}{3}e^{-(\sigma_{\mathrm{S}}t)^{2}}+\frac{1}{3}\right]+(1-f_{\mathrm{S}})e^{-\lambda_{\mathrm{P}}t},
\end{equation}
where $\sigma_{\mathrm{S}}$ and $\lambda_{\mathrm{P}}$ stand for the fast static and slow dynamic relaxation rates due to frozen moments and to the remaining paramagnetic-like moments respectively. The temperature evolution of $f_{\mathrm{S}}$ (see Figure~\ref{fig6}(a)) reveals a broad transition which starts around $14$~K, coinciding with the steep rise in the dc susceptibility~\cite{Sun2016}, towards a regime in which nearly $62\pm5$~\% of the muon spins experience a frozen field below about $5$~K. As shown in Figure~\ref{fig6}(c), $\lambda_{\mathrm{P}}$ levels off around the small value of $0.062\pm0.005$~$\mu$s$^{-1}$ in this regime. We note that $f_{\mathrm{S}}$ and $\lambda_{\mathrm{P}}$ evolve in the same way in temperature, indicating that these two responses at around $14$~K are intrinsically correlated and that none of them is caused by an extrinsic impurity phase. The ground state in the $x=0$ compound therefore appears to be inhomogeneous with a larger proportion of muon spins that detect a static local field.

\begin{figure}[h]\centering
\includegraphics[width=\linewidth]{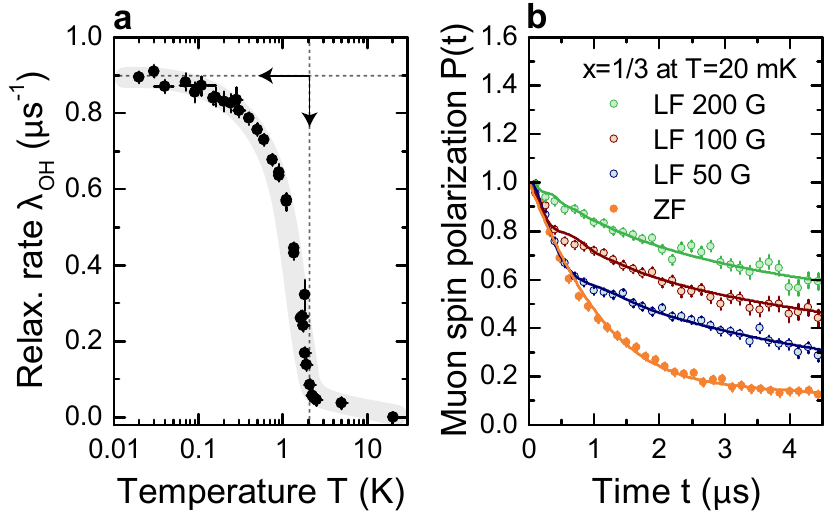}
\caption{\label{fig7}(a) Temperature evolution of the dynamical relaxation rate $\lambda_{\mathrm{OH}}$ extracted from fits of the zero field $x=1/3$ data to Equation~\ref{eq3}. The light-gray line is a guide to the eye. The downward arrow indicates the position of the anomaly in specific heat and the leftward arrow indicates the estimated saturating value of $\lambda_{\mathrm{OH}}$ as $T\rightarrow0$. (b) Decoupling experiment at base temperature ($20$~mK) on the $x=1/3$ compound for longitudinal applied fields from $0$ to $20$~mT. The solid lines are fits according to Equation~\ref{eq3} and Equation~\ref{eq4} (see text).}
\end{figure}

\indent For the $x=1/3$ compound, in order to track the weak temperature evolution of the electronic relaxation in zero applied field, we used the following modified version of Equation~\ref{eq1}:
\begin{equation}
\label{eq3}
P_{\mathrm{ZF}}(t)=fP^{\mathrm{OH}}(t)e^{-\lambda_{\mathrm{OH}}t}+(1-f)\mathrm{KT}_{\Delta}(t)e^{-\lambda t},
\end{equation}
which includes now the dynamical relaxation of the copper electronic spins felt by the muon spins at the two types of stopping sites through the relaxation rates $\lambda_{\mathrm{OH}}$ and $\lambda$. In this model, we could fit the zero field data on the whole temperature range with the exponential rates as only adjustable parameters while all others are fixed to their high temperature values. We obtain $\lambda\sim0.05(1)$~$\mu$s$^{-1}$ at all temperatures, which suggests that the muons that are not in a $\mu$-OH complex are too far from the kagome planes to be sensitive to the spin dynamics. All the temperature evolution is thus accounted by $\lambda_{\mathrm{OH}}$ as depicted in Figure~\ref{fig7}(a). Above $\sim5$~K, the copper electronic spins are in the fast fluctuating paramagnetic regime. Upon cooling below $\sim3$~K, the electronic spin dynamics strongly slows down in the same range as the peak observed in the specific heat~\cite{Puphal2017}, then levelling off gradually. Slow $T$-independent fluctuations persist down to the lowest temperature $T=20$~mK.

\indent Next, we analyse the effect of an applied longitudinal field at base temperature, which provides a more detailed understanding of the origin of the relaxation in the ground state. As seen in Figure~\ref{fig7}(b), when the applied field increases, two distinct relaxations are observed: an initial fast relaxation is progressively suppressed while a slower one is almost unaffected. We thus used the following two-component expression:
\begin{equation}\label{eq4}
P_{\mathrm{LF}}(t)=f_{1}\mathrm{DGKT}_{1}(t)+f_{2}\mathrm{DGKT}_{2}(t),
\end{equation}
where DGKT$_{i}$ is the dynamical Kubo-Toyabe function modeling the relaxation due to a gaussian field distribution of width $\Delta_{i}/\gamma_{\mu}$ fluctuating at the rate $\nu_{i}$. By fitting globally the data corresponding to three different applied fields in Figure~\ref{fig7}(b), we find that a fraction $f_{1}\sim35.5\pm3.7$~\% of muons probes a distribution $\Delta_{1}/\gamma_\mu\sim2.8(3)$~mT of fluctuating fields with $\nu_{1}\sim3.7(4)$~$\mu$s$^{-1} > \Delta_{1}$ and that a fraction $f_{2}=1-f_{1}$ of muons probes a distribution $\Delta_{2}/\gamma_{\mu}\sim1.4(2)$~mT of slower, nearly static, fluctuating fields with $\nu_{2}\sim1.1(1)$~$\mu$s$^{-1}$~$\sim\Delta_{2}$. These two contributions cannot be attributed to the two muon stopping sites because, in this case, the larger $\Delta_{1}$ should be associated to the $\mu$-OH site fraction but $f_{1}$ does not match $f$. Our results rather point to a spatial inhomogeneity of the fluctuation rate. Note that Equation~\ref{eq2} is a simplified version of Equation~\ref{eq4} in the case of one fully static and one fully dynamical components, showing a similar inhomogeneity in the ground states of the two compounds.

\section{Discussion}

\begin{figure}\centering
\includegraphics[width=\linewidth]{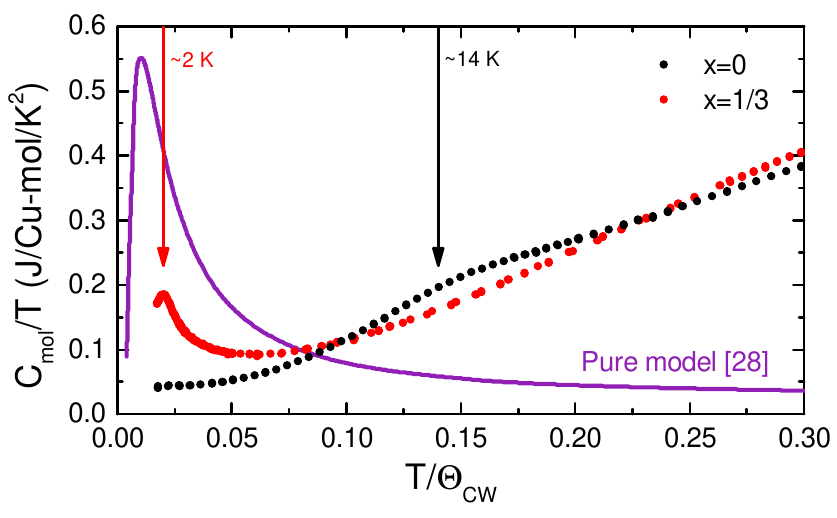}
\caption{\label{fig8}Comparison of the specific heat measured on the $x=0$ and $x=1/3$ compounds with the result for the pure antiferromagnetic (KHAF) model on the kagome lattice (adapted from Reference~\cite{Bernu2015} with $J=\theta_{CW}=100$~K). In the absence of a reliable model for the phononic contribution, the latter was not subtracted in the $x=0$ and $x=1/3$ data. The arrows indicate the positions of the anomalies discussed in the text.}
\end{figure}

In summary, the $x=1/3$ compound does not exhibit static magnetism in its ground state but persistent dynamics, the key feature of a QSL. The establishment of this state is concomitant with a strong slowing down of the spin fluctuations around $2$~K, which coincides with the peak already observed in specific heat (see Reference~\cite{Puphal2017} and Figure~\ref{fig8}). On the contrary, the $x=0$ compound experiences a magnetic transition around $14$~K although only $\sim62$~\% of the muon spins probe glassy static magnetism while the remaining $\sim38$~\% probe a correlated paramagnetic-like regime with slow persistent fluctuations. The development of this unusual heterogeneous phase is in accordance with the reported features in dc susceptibility~\cite{Sun2016} and a broad hump observed in heat capacity (Figure~\ref{fig8}).\\
\indent These two distinct ground states necessarily stem from the few structural differences between the two compounds (see Figure~\ref{fig1}). In Figure~\ref{fig8}, the specific heat curves for the $x=0$ and $x=1/3$ compounds are also compared to the theoretical prediction for the pure kagome model (KHAF), namely the first neighbor antiferromagnetic Heisenberg hamiltonian on the kagome lattice~\cite{Bernu2015}. In the absence of a reliable model for the phononic contribution, the latter was not subtracted from the experimental data which explains the large discrepancy with the theoretical curve (magnetic contribution only) at high temperature ($T/\theta_{CW}\gtrsim0.05$). The anomaly at about $2$~K in the specific heat of the $x=1/3$ compound might be reminiscent of the large maximum expected at low $T$ in the pure model. But, most noticeably, the release of entropy at $2$~K and $14$~K in the $x=1/3$ and $x=0$ compounds is surprisingly small, suggesting that in both compounds the correlations build up very progressively below $\theta$ at variance with the pure model where the release of entropy is shifted characteristically downward to low $T\ll\theta$. This comparison suggests that the magnetic models for both compounds are substantially modified as compared to the ideal kagome one.\\
\indent We first discuss the rather unexpected occurrence of a magnetic transition in the \mata\ ($x=0$) compound. Indeed, this strongly two-dimensional material features a perfect kagome lattice with a single copper site (the triangles are equilateral). The large antiferromagnetic Weiss temperature of about $-100$~K suggests a dominant first nearest neighbor antiferromagnetic interaction $J_1$. And indeed, the corresponding Cu-O-Cu superexchange path  forms an angle of $115.69$\textdegree, quite similar to the one in herbertsmithite where $J_{1}\sim-180$~K and well above the reduced value of $\sim105$\textdegree\ in kapellasite which produces a weak ferromagnetic interaction. From these arguments, a natural  minimal magnetic model would be the KHAF, however this model results in a QSL ground state.\\
\indent Additional interactions must be relevant to induce the magnetic transition. A first possibility is the antisymmetric Dzyaloshinskii-Moriya (DM) interaction which is allowed by the low symmetry of kagome lattice. The out-of-plane DM component $D$ is known to drive a quantum phase transition towards the uniform $q=0$ N\'{e}el state~\cite{Elhajal2002,Messio2010}. The critical point was first estimated to $D_{c}/J\sim0.1$~\cite{Cepas2008} but could maybe scale down to $D_{c}/J=0.012(2)$ according to a recent numerical approach~\cite{Lee2018}. In any case, $D_{c}$ lies in a realistic range for Cu based kagome materials~\cite{Zorko2008,Zorko2013}. The second order nature of this quantum transition would give a natural explanation for the apparently reduced moment $\sim0.1\mu_{\mathrm{B}}\propto(D-D_{c})^{\alpha}$ ($1/2\lesssim\alpha\lesssim1$) observed in the magnetic phase. The determination of the DM interaction, for instance by Electron Spin Resonance (ESR), would help clarifying this issue. A second possibility is the competition with further neighbor interactions. The quantum ternary phase diagram $J_{1}$-$J_{2}$-$J_{\mathrm{d}}$, where $J_{2}$ and $J_{\mathrm{d}}$ stand for the couplings between second nearest neighbors and further neighbors along the diagonals of the hexagons, has been determined for the case where all interactions are antiferromagnetic~\cite{Bieri2016}. The ground state is found to be long range ordered ($q=0$ state) only for sizeable $J_{2}/(J_{1}+J_{2}+J_{\mathrm{d}})\gtrsim0.27$, \emph{i.e.} $J_{2}\sim27$~K if $J_{\mathrm{d}}\sim0$. Although it seems unlikely that $J_{2}$ could be that strong, first principle calculations are clearly desirable to determine the relevance and signs of the further neighbor couplings.\\
\indent Last, we note that all these perturbations would drive the kagome model to a long range ordered $q=0$ ground state at variance with the observation of an only partially static and mostly disordered magnetic ground state. Disorder could be invoked to explain this discrepancy but here, at variance with kapellasite, no chemical disorder between the magnetic Cu$^{2+}$ and the diamagnetic Y$^{3+}$ was detected. We recall nonetheless that the coexistence of small frozen moments with dynamically fluctuating spins is very similar to what was initially found in powder samples of vesignieite \CuBaVOH\ \cite{Colman2011,Quilliam2011}, while it was eventually shown that all spins order at low-temperature in single crystals~\cite{Yoshida2013,Boldrin2018}. It seems thus that even minute disorder, such as non-optimal crystallization, may impede the full development of long range order in these frustrated magnets.\\
\indent We now focus on \matb\ ($x=1/3$) where the establishment of a QSL is also rather counterintuitive. Here, the magnetic Cu$^{2+}$ form an interesting variation of the kagome lattice with, for each triangle two copper sites Cu$1$, Cu$2$ and three distances labelled $d1$, $d2$, $d3$ in Figure~\ref{fig1}. The corresponding superexchange path angles are respectively Cu$1$-O$2$-Cu$1$ $\sim117.18$\textdegree, Cu$1$-O$4$-Cu$2$ $\sim114.80$\textdegree\ and Cu$1$-O$3$-Cu$2$ $\sim111.51$\textdegree giving three different, likely all antiferromagnetic, magnetic interactions on the three bonds of each triangle, with the largest magnitude for bond $1$ and the smallest for bond $3$. The triangles are related by $6$ fold symmetry around the hexagons consisting of Cu$1$ ions only. To our knowledge this model has not been worked out theoretically and it is thus not clear if a QSL ground state is robust to the anisotropy. On general grounds, the anisotropy reduces drastically the classical degeneracy of the ground state and suppresses the local zero-energy modes specific to the isotropic model. However, in the distinct but also spatially anisotropic $J-J^{'}$ model, some subextensive degeneracy subsists and is robust to thermal and quantum fluctuations~\cite{Wang2007,Yavorskii2007}.\\
\indent Another interesting open issue is the effect of DM interaction, which should be similar to the one in the $x=0$ case, in this modified kagome model. Additionally, at variance with the kapellasite structure, some yttrium ions are displaced from the center of the hexagons along the $\mathbf{c}$ direction yielding a buckling of the Cu-Y planes as shown in Figure~\ref{fig1}(b). It is conceivable that the diagonal superexchange $J_{\mathrm{d}}$ which involves the Y orbitals is decreased compared to the case of $x=0$. Here again first principle calculations would be helpful to validate the former nearest neighbor anisotropic model as a good starting description for the $x=1/3$ compound and to quantify the interactions.\\

\section{Conclusion}

The ongoing efforts for doping a QSL remain unsuccessful so far but gave birth to two new systems which both question our understanding of the deviations to the pure antiferromagnetic Heisenberg model on the kagome lattice. Namely, the robustness of the QSL ground state for kagome system to the allowed DM interaction is still debated and relevant for all kagome materials. The dynamical, liquid-like, ground state in the $x=1/3$ compound offers a new anisotropic magnetic model for which the ground state and its connection to the isotropic case remain to be elucidated. In this line, first-principles calculations and local probe ESR and NMR studies will be tools of choice to gain insight into the magnetic hamiltonians of the $x=0$ and $x=1/3$ compounds.

\section*{Acknowledgments}

This work is partly based on experiments performed at the Swiss spallation neutron source SINQ and the Swiss Muon Source S$\mu$S, Paul Scherrer Institute, Villigen, Switzerland.
This research project has been supported by the French Agence Nationale de la
Recherche under Grant ANR-18-CE30-0022 ``LINK'' and by the German Science Foundation (DFG) through grant SFB/TR49.

\end{document}